\documentclass[
journal=nalefd,
manuscript=letter,
]{achemso}

\usepackage[version=3]{mhchem}
\usepackage{amsmath}
\usepackage[separate-uncertainty=true]{siunitx} 

\newcommand{\MBI}{Max Born Institute for Nonlinear Optics and Short Pulse Spectroscopy, 12489 Berlin, Germany}
\newcommand{\FBH}{\small Ferdinand-Braun-Institut gGmbH, Leibniz-Institut für Höchstfrequenztechnik, 12489 Berlin, Germany}
\newcommand{\HZB}{\small Helmholtz-Zentrum für Materialien und Energie GmbH, 14109 Berlin, Germany}
\newcommand{\ZELMI}{\small Technische Universität Berlin, Zentraleinrichtung Elektronenmikroskopie (ZELMI), 10623 Berlin, Germany}
\newcommand{\IOAP}{\small Technische Universität Berlin, Institut für Optik und Atomare Physik, 10623 Berlin, Germany}
\newcommand{\DESY}{\small Deutsches Elektronen-Synchrotron (DESY), 22607 Hamburg, Germany}
\newcommand{\RU}{\small Radboud University, Institute for Molecules and Materials (IMM), 6525 AJ Nijmegen, Netherlands}
\newcommand{\MPIIS}{\small Max Planck Institute for Intelligent Systems, 70569 Stuttgart, Germany}

\author{Lisa-Marie Kern}
\affiliation{\MBI}
\author{Bastian Pfau}
\affiliation{\MBI}
\email{*bastian.pfau@mbi-berlin.de}
\author{Victor Deinhart}
\affiliation{\MBI}
\alsoaffiliation{\FBH}
\alsoaffiliation{\HZB}
\author{Michael Schneider}
\affiliation{\MBI}
\author{Christopher Klose}
\affiliation{\MBI}
\author{Kathinka Gerlinger}
\affiliation{\MBI}
\author{Steffen Wittrock}
\affiliation{\MBI}
\author{Dieter Engel}
\affiliation{\MBI}
\author{Ingo Will}
\affiliation{\MBI}
\author{Christian M.\ Günther}
\affiliation{\ZELMI}
\author{Rein Liefferink}
\affiliation{\RU}
\author{Johan H.\ Mentink}
\affiliation{\RU}
\author{Sebastian Wintz}
\affiliation{\MPIIS}
\author{Markus Weigand}
\affiliation{\HZB}
\author{Meng-Jie Huang}
\affiliation{\DESY}
\author{Riccardo Battistelli}
\affiliation{\HZB}
\author{Daniel Metternich}
\affiliation{\HZB}
\author{Felix Büttner}
\affiliation{\HZB}
\author{Katja Höflich}
\affiliation{\FBH}
\alsoaffiliation{\HZB}
\author{Stefan Eisebitt}
\affiliation{\MBI}
\alsoaffiliation{\IOAP}

\title{Deterministic Generation and Guided Motion of Magnetic Skyrmions by Focused He$^+$-Ion Irradiation}


\begin{document}

\begin{tocentry}
\includegraphics{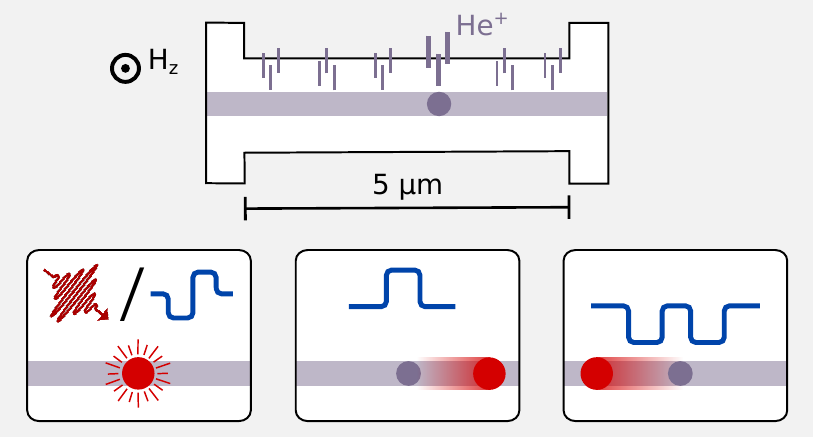}
For Table of Contents Only
\end{tocentry}

\begin{abstract}
Magnetic skyrmions are quasiparticles with non-trivial topology, envisioned to play a key role in next-generation data technology while simultaneously attracting fundamental research interest due to their emerging topological charge. In chiral magnetic multilayers, current-generated spin--orbit torques or ultrafast laser excitation can be used to nucleate isolated skyrmions on a picosecond timescale. Both methods, however, produce randomly arranged skyrmions, which inherently limits the precision on the location at which the skyrmions are nucleated. Here, we show that nanopatterning of the anisotropy landscape with a He$^+$-ion beam creates well-defined skyrmion nucleation sites, thereby transforming the skyrmion localization into a deterministic process. This approach allows to realize control of individual skyrmion nucleation as well as guided skyrmion motion with nanometer-scale precision, which is pivotal for both future fundamental studies of skyrmion dynamics and applications.
\end{abstract}

Magnetic skyrmions are topological quasiparticles that can be as small as a few nanometers \cite{fert2017magnetic,roessler2006spontaneous,caretta2018fast}. During the last decade, research on magnetic skyrmions attracted interest from scientific \cite{fert2013skyrmions,tomasello2014strategy} and industrial \cite{luo2021skyrmion} research communities due to the skyrmion’s fascinating properties emerging from its topological charge. They can exist in thin film materials with perpendicular magnetic anisotropy (PMA) and are stabilized by stray fields and the Dzyaloshinskii-Moriya interaction (DMI) \cite{yu2012magnetic,muhlbauer2009skyrmion,nagaosa2013topological, romming2013writing,buttner2015dynamics,woo2016observation,moreau2016additive}. In particular, skyrmions in chiral magnetic multilayer systems can occur as isolated particle-like textures at room temperature and in a low external magnetic field regime down to zero field \cite{moreau2016additive,legrand2017room,buttner2018theory}. 

Great advances have been reported in generating, annihilating and shifting skyrmions via spin--orbit torques (SOT) induced by electric currents inside a suitable magnetic racetrack \cite{woo2016observation,buttner2017field,yu2017room,legrand2017room,caretta2018fast,woo2018deterministic,litzius2017skyrmion,jiang2017direct}. Moreover, recent studies have revealed that faster and potentially more energy-efficient skyrmion generation is feasible by replacing current pulses with femtosecond laser pulses, allowing optical nucleation even with a single laser pulse \cite{je2018creation,finazzi2013laser,buttner2021observation}. 

While the underlying mechanisms are different, both current-induced and optical nucleation suffer from a randomness in the spatial distribution of the skyrmions nucleated. SOT magnetic switching and nucleation of a skyrmion are mediated either by an applied symmetry-breaking in-plane magnetic field \cite{liu2012spin,garello2014ultrafast} or by a locally modified anisotropy at a magnetic defect site in combination with DMI \cite{buttner2017field,woo2018deterministic}. Natural magnetic defects caused by structural and chemical inhomogeneity can act as such nucleation sites, but they also influence the SOT-driven skyrmion motion \cite{legrand2017room, reichhardt2021statics}. Since the density and distribution of these defects depend on the growth process of the material, the localization of the skyrmions remains poorly controllable. In case of the optical nucleation, skyrmions emerge from spin fluctuations during a high-temperature phase after the laser excitation. They finally appear randomly distributed inside the illuminated area of the magnetic film \cite{buttner2021observation, gerlinger2021application}. However, a controllable, reproducible and reliable localization of the skyrmion nucleation is often required---both for future applications in data technology \cite{zhang2020skyrmion, finocchio2021promise} as well as for fundamental research, e.g., to realize repetitive pump--probe experiments on skyrmion dynamics \cite{woo2018deterministic}. 

Previous attempts to control the localization of skyrmions include structural patterning of the magnetic film with notches \cite{bedau2007angular,buttner2017field,finizio2019deterministic} and discs \cite{buttner2015dynamics} as well as the introduction of defects \cite{liu2013mechanism,hanneken2016pinning} and nano-pockets \cite{pathak2021geometrically} to define positions for skyrmion nucleation. However, all of these methods include significant structural or even geometrical modifications of the magnetic racetrack and render an unhampered motion of the skyrmion after creation difficult. Recent work also employed Ga$^+$-ion bombardment to create artificial defects as pinning sites for skyrmions \cite{fallon2020controlled,deJong2021local}. Due to the strong structural impact of these heavy ions, it again remains an open question if a controlled detachment of the skyrmion from the pinning site created in this fashion is possible. In contrast, nanopatterning with a focused He$^+$-ion beam is suitable to prepare magnetic channels for magnetic domain and skyrmion motion \cite{juge2021helium}. Light-ion irradiation causes only minor structural reorganization on the atomic scale, giving rise to an increased interfacial roughness and intermixing of layers---without affecting the topography \cite{chappert1998planar,devolder2000light,fassbender2004tailoring,fassbender2008magnetic,zhang2020ion,dunne2020helium,an2021improved}. As a consequence, magnetic properties such as coercivity, PMA and DMI can be controlled at the nanoscale \cite{fassbender2004tailoring,fassbender2008magnetic,dunne2020helium,nembach2020tuning,krupinski2021control}. 

In this work, we use a focused He$^+$-ion beam to control the localization of current-induced and laser-induced magnetic skyrmions in ion-irradiated areas with different shapes and sizes. In conjunction with the applied magnetic field, we can turn the nucleation into a fully deterministic process---guaranteeing skyrmions to be generated in irradiated sites while simultaneously suppressing their creation in non-irradiated regions. Furthermore, we demonstrate controlled SOT-induced detachment from such a nucleation site and the subsequent guided motion of a single skyrmion over several micrometers along a linear path, precisely defined by ion irradiation, thereby effectively diminishing the transverse skyrmion drift due to the skyrmion Hall effect. Our results show the large potential of magnetic anisotropy patterning by He$^+$ irradiation in multilayer structures for applied and fundamental research on isolated skyrmions. 

\begin{figure}[htbp]
	\begin{center}
		\includegraphics{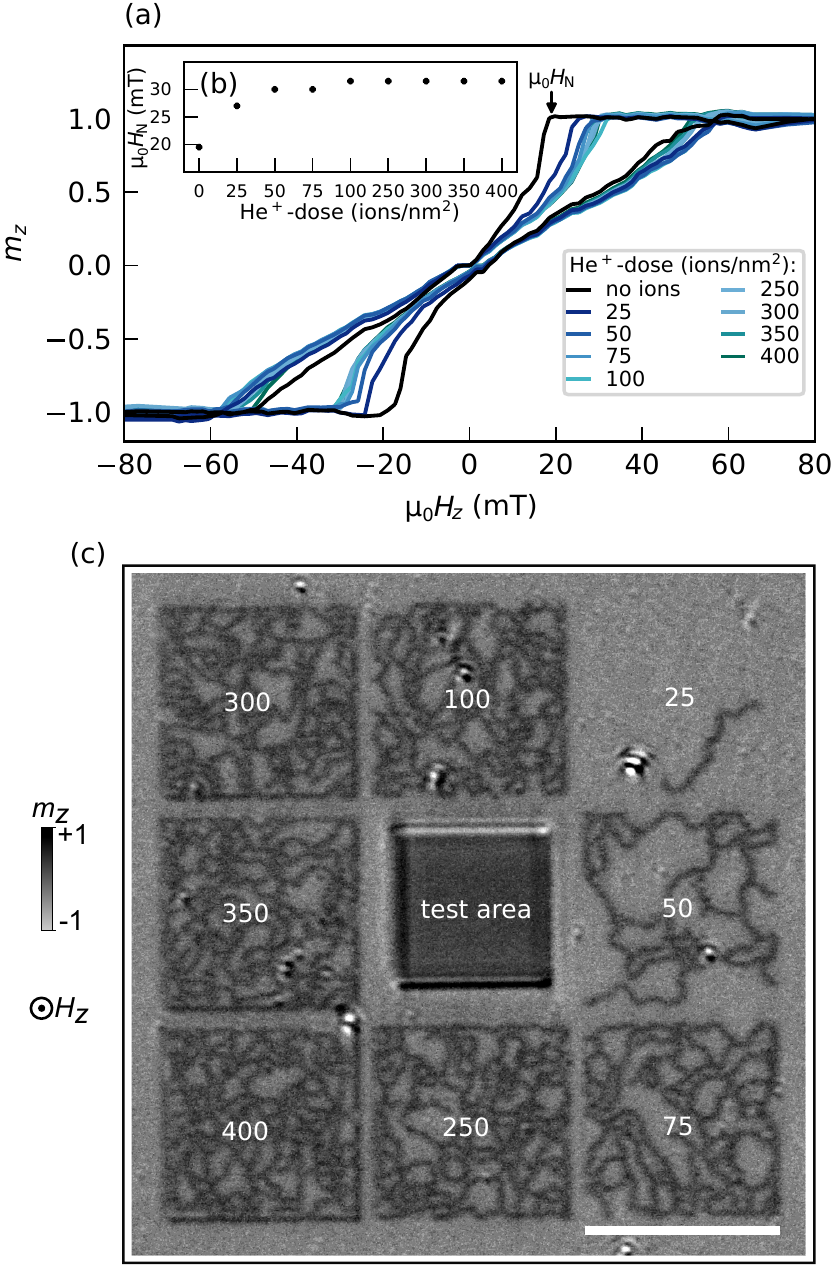}
	\end{center}
	\caption{\label{fig:hysteresis}\textbf{Magnetic hysteresis of He$^+$-ion irradiated multilayer.} (a) Normalized out-of-plane magnetization $m_z$ as a function of magnetic field $H_z$ for variable He$^+$-ion dose. (b) Nucleation field $H_{\mathrm{N}}$ as a function of applied He$^+$-ion dose. (c) Kerr image of irradiated areas in the magnetic film for $\mu_0 H_z = \SI{26}{mT}$ after saturation. Ion doses in ions/nm$^2$. Scalebar corresponds to \SI{10}{\micro m}.}
\end{figure}

We prepared ferromagnetic multilayers with a nominal composition of Ta(\SI{3}{nm})/\newline Pt(\SI{4}{nm})/[Pt(\SI{2.5}{nm}/Co$_{60}$Fe$_{25}$B$_{15}$(\SI{0.95}{nm})/MgO(\SI{1.4}{nm})]$_{15}$/Pt(\SI{2}{nm}) (see Supporting information for more details on sample fabrication). The material system is well-known from previous experiments on skyrmion nucleation and supports both SOT- and laser-assisted nucleation \cite{buttner2017field, buttner2021observation, gerlinger2021application}. He$^+$-ion nanopatterning was carried out with a ZEISS Orion Nanofab (acceleration voltage: \SI{30}{kV}). We locally irradiated the magnetic film with He$^+$-ion doses between \num{25} and \num{400} ions/nm$^2$ in predefined patterns with a nominal resolution below \SI{10}{nm}. However, the ion impact diameter broadens to approximately \SI{20}{nm} (full width at half maximum, FWHM) due to the ion collision cascade in our more than \SI{80}{nm} thick multilayer stack (see Supporting Information for details). In Fig.~\ref{fig:hysteresis}a, we present the magnetic hysteresis, measured with Kerr microscopy on the pristine film and the irradiated regions as shown in Fig.~\ref{fig:hysteresis}c. In the irradiated regions, at doses below \SI{50}{ions \per \nano\metre\squared}, the field $H_{\mathrm{N}}$ allowing for spontaneous nucleation increases with increasing dose. Above \SI{50}{ions \per \nano\metre\squared}, this dependence diminishes and no further increase of $H_{\mathrm{N}}$ is observed for ion doses above \SI{100}{ions \per \nano\metre\squared} (see also Fig.~\ref{fig:hysteresis}b). This behavior is in line with stopping and range of ions in matter (SRIM) simulations for the parameters in our experiment: The additional intermixing effect caused by the He$^+$-ions decreases with increasing ion dose and saturates for ion doses above \num{100} ions/nm$^2$ (see Supporting Information). The characteristic out-of-plane hysteresis shape and the presence of labyrinth-like domains confirm that the magnetic multilayer retains its perpendicular anisotropy and supports magnetic domains in the nanometer range for all He$^+$-ion doses used here. 

\begin{figure}[htbp]
	\begin{center}
		\includegraphics{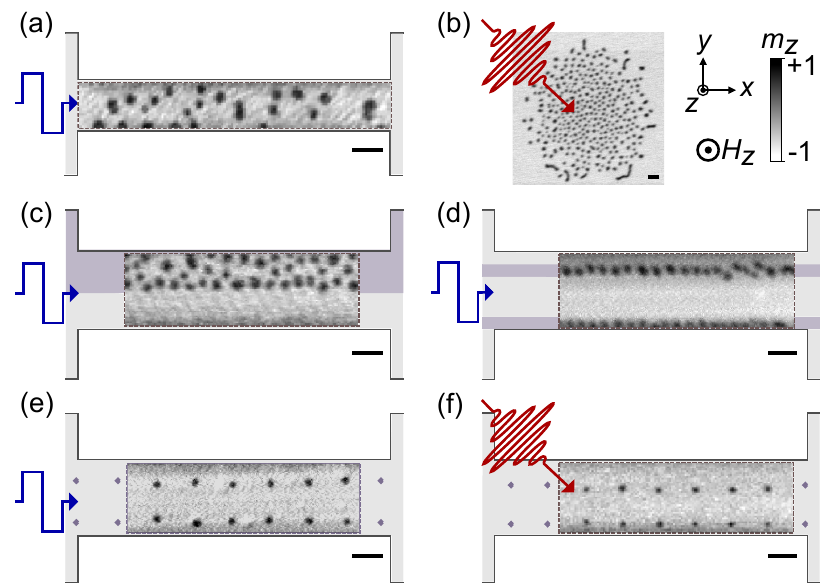}
	\end{center}
	\caption{\label{fig:scheme}\textbf{Localized current-induced and optical skyrmion nucleation.} (a--b) Stochastic nucleation in the pristine material either by a current pulse (a, current density $j_x = \SI{\pm 1.1e12}{A \per m \squared}$, $\mu_0 H_z = \SI{26}{mT}$) or  by an optical laser pulse (b, peak fluence \SI{53}{mJ/cm \squared}, $\mu_0 H_z = \SI{40}{mT}$). In (c--f), a scheme of the structure indicating the He$^+$ irradiated regions (red) has been overlayed with images of the out-of-plane magnetization obtained by STXM. (c--d) Current-induced nucleation in an irradiated area (c, \num{100} ions/nm$^2$ dose, $j_x = \SI{\pm 1.3e12}{A \per m \squared}$, $\mu_0 H_z = \SI{27}{mT}$) and on lines (d, \SI{200}{nm} width, \num{350} ions/nm$^2$ dose, $j_x = \SI{\pm 7.7e11}{A \per m^2}$, $\mu_0 H_z =\SI{40}{mT}$). (e) Current-induced nucleation in a dot array of dose \num{350} ions/nm$^2$ ($j_x$ as in (d), $\mu_0 H_z =\SI{47}{mT}$). (f) Optical nucleation in a dot array of dose \num{350} ions/nm$^2$ (peak fluence $\SI{45}{mJ/cm \squared}$, $\mu_0 H_z = \SI{51.5}{mT}$). (a--f) Skyrmions are created from a \emph{single} current or laser pulse. Scalebars correspond to \SI{500}{nm}.}
\end{figure}

First, we present skyrmion nucleation in \num{2}D-, \num{1}D- and \num{0}D-like ion-irradiated areas with homogeneous ion doses within each pattern. We then turn our attention to the movement of skyrmions in more complex irradiation profiles. Both the SOT and laser-induced skyrmion nucleation were investigated in samples patterned into a racetrack geometry \cite{fert2013skyrmions, yu2017room} (Fig.~\ref{fig:scheme}), irradiated with He$^+$-ions in different shapes and doses. We imaged the magnetic textures nucleated via scanning transmission x-ray microscopy (STXM) and Fourier-transform holography (FTH) in transmission, utilizing x-ray magnetic circular dichroism (XMCD) to obtain sensitivity to the magnetization perpendicular to the sample plane ($m_z$) (see Supporting Information for more experimental details). 

For the nucleation experiments, we first fully saturated our magnetic film in an external magnetic field. In line with the modification of the hysteresis behavior for different irradiation doses, we reduced the magnetic field to a value for which spontaneous nucleation of stripe domains remains suppressed but skyrmion nucleation from an external stimulus is possible \cite{buttner2017field,gerlinger2021application}. At the outset, we confirmed that our non-irradiated Pt/CoFeB/MgO multilayer supports skyrmion nucleation. As a reference for the unmodified magnetic film in Fig.~\ref{fig:scheme}a and b, we present images of the skyrmion patterns created by applying a single bipolar current pulse ($j_x = \SI{\pm 1.1e12}{A \per m \squared}$) and a single infra-red (IR) laser pulse (\SI{53}{mJ/cm \squared}), respectively (see Supporting Information for pulse parameters). For the non-irradiated magnetic film, we observed a random distribution of skyrmions in both cases. The reference sample in Fig.~\ref{fig:scheme}b was not patterned into a racetrack and skyrmions form in the entire area illuminated by the laser where the laser fluence overcomes the nucleation threshold \cite{buttner2021observation,gerlinger2021application}. In samples with an ion irradiated \num{2}D area covering half the racetrack (c) and with a \num{1}D line pattern (d), we applied a single bipolar current pulse to create skyrmions ((c) $j_x = \SI{\pm 1.3e12}{A \per m \squared}$, (d) $j_x = \SI{\pm 7.7e11}{A \per m^2}$). To suppress nucleation in non-irradiated regions, we increased the applied field compared to the nucleation in non-irradiated racetracks. In this fashion, skyrmion nucleation only occurs in irradiated regions. Within these extended \num{2}D and \num{1}D ion-irradiated areas, the skyrmions preserve their spherical shape compared to the pristine film and appear almost homogeneously distributed. In Fig.~\ref{fig:scheme}e and f, we show the results from current-induced and optical excitations in racetracks with predefined \num{0}D dot patterns (\SI{100}{nm} diameter) by He$^+$-ion irradiation ((e) $j_x = \SI{\pm 7.7e11}{A \per m^2}$, (f) $\SI{45}{mJ/cm \squared}$). For appropriate external fields, we observe skyrmion nucleation exclusively and exactly in the predefined dots, with the skyrmion diameter of about \SI{125}{nm} slightly exceeding the size of the He$^+$ patterned dots. 

\begin{figure}[htbp]
	\begin{center}
		\includegraphics{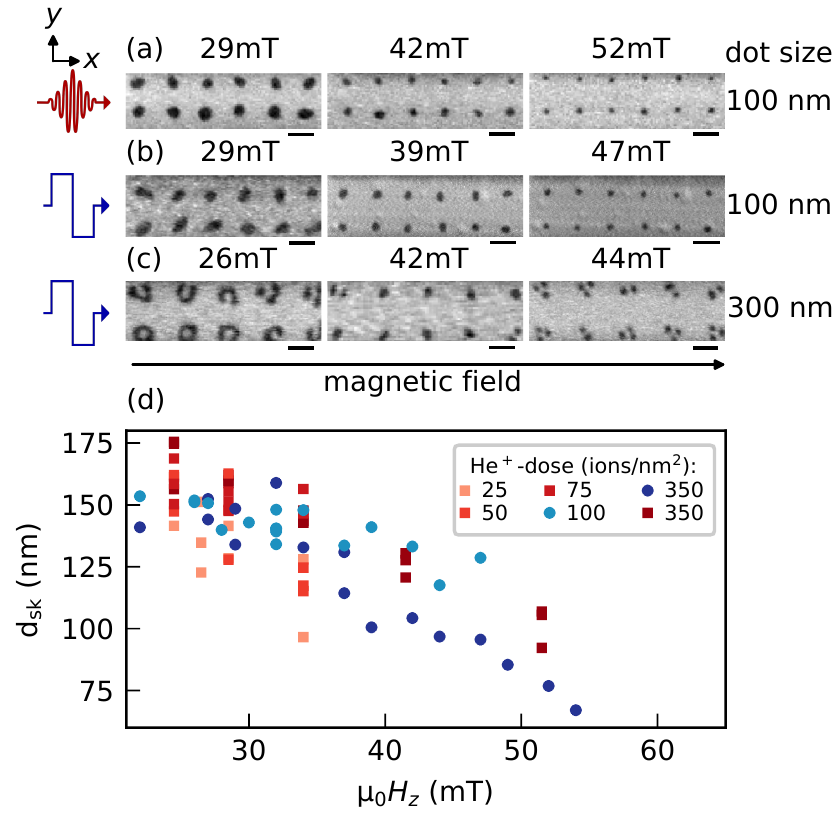}
	\end{center}
	\caption{\label{fig:statistics}\textbf{Magnetic field impact on skyrmion diameter and nucleation reliability.} (a--c) STXM images after skyrmion nucleation in magnetic racetracks with He$^+$ patterned dot arrays (\num{350} ions/nm$^2$ dose). Nucleation method (symbols on the left), applied field (above each image), and irradiated dot diameter (right) are indicated. Scalebars correspond to \SI{500}{nm}. (d) Skyrmion diameter $d_\mathrm{sk}$ as a function of the applied field in magnetic magnetic racetracks with He$^+$ patterned arrays of dots with \SI{100}{nm} diameter. Data points with red color shades and squared markers correspond to laser-induced nucleation, blue shades and round markers to current-induced nucleation.}
\end{figure}

In Fig.~\ref{fig:hysteresis} and Fig.~\ref{fig:scheme}, we demonstrated that we can employ the applied field to effectively suppress nucleation in non-irradiated regions. Next, we discuss the influence of the magnetic field on the skyrmion formation inside the predefined regions. For He$^+$ patterned dots of \SI{100}{nm} diameter, we observed reliable nucleation at the dots by both laser pulses and SOT pulses, as shown in Fig.~\ref{fig:statistics}a and b, respectively. Skyrmions nucleate at all dots, and exclusively there, over a broad range of fields applied. Similar to skyrmions in homogeneous media, the skyrmion diameter continuously decreases with increasing magnetic field. The skyrmion size is, thus, not fixed to the size of the He$^+$ irradiated area. 
In contrast, for the skyrmion nucleation at He$^+$ patterned dots of \SI{300}{nm} diameter (Fig.~\ref{fig:statistics}c), we observe the generation of non-compact magnetization textures. At low magnetic field, a single bipolar current pulse creates a variety of almost closed or fully closed ring-shaped domains (so-called skyrmionia \cite{gobel2021beyond} or target skyrmions \cite{kent2019generation}) surrounding the irradiated dot. Increasing the magnetic field mostly leads to the formation of a single circular skyrmion per irradiated dot. At even higher applied fields, several magnetic skyrmions (mostly three) stabilize inside the He$^+$-ion irradiated areas. We infer that the most suitable choice for controlled nucleation of individual skyrmions is to rely on He$^+$-ion irradiated dots of a size comparable to or smaller than the skyrmion diameter supported by the material at a particular field. Nevertheless, larger He$^+$ dots provide an interesting route to create and investigate topological textures beyond charge-unity skyrmions \cite{gobel2021beyond}.

In Fig.~\ref{fig:statistics}d, we present the skyrmion diameter on the array of \SI{100}{nm} sized He$^+$-ion irradiated dots  for different applied ion doses in a range from \num{25} to \num{350} ions/nm$^2$ and as a function of the external field. The current and optical excitation amplitudes were chosen such that for the lowest magnetic field investigated, all predefined dots were filled with skyrmions while nucleation in non-irradiated regions remains suppressed. The excitation setting per sample was kept constant while varying the magnetic field. The skyrmion diameter was determined from the full width at half maximum (FWHM) of the central autocorrelation peak of the STXM images as FWHM/$\sqrt2$ to account for the skyrmions' self-convolution. Each data point thus corresponds to the average diameter of several skyrmions recorded in one STXM image. In Fig.~\ref{fig:statistics}d, we observe a decrease in skyrmion diameter with increasing field in the direction opposite to the net perpendicular magnetization of the skyrmion from about \SI{155}{nm} at \SI{25}{mT} to \SI{75}{nm} at \SI{52}{mT}. We attribute the large scatter to a spatial variation of anisotropy and DMI and a variation of the location of the irradiated regions within the narrowed racetrack affecting stray fields \cite{juge2018magnetic}. Within this variance, there is no apparent systematic dependence of skyrmion diameter on ion dose. 

\begin{figure*}[htbp]
	\begin{center}
		\includegraphics[width=\linewidth]{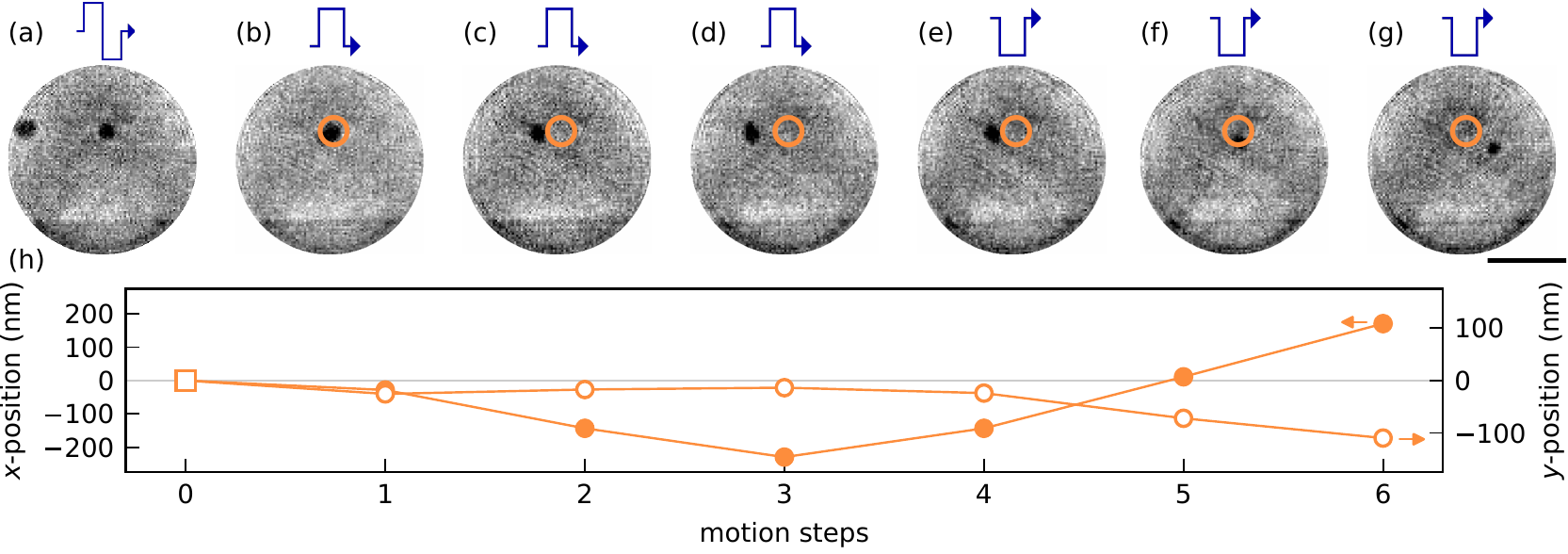}
	\end{center}
 	\caption{\label{fig:detachment}\textbf{Controlled detachment from a He$^+$-ion irradiated dot.} (a--g) X-ray holography images of a skyrmion detaching from and moving over its nucleation site due to single current pulses at constant applied field of $\mu_0 H_z = \SI{36}{mT}$. (a) Current-induced nucleation at predefined dots (\num{50} ions/nm$^2$ dose, $j_x = \SI{\pm 1.4e12}{A \per m^2}$), (b--g) detachment and free motion ($j_x = \SI{4.3e11}{A \per m^2}$). The orange circle marks the position of the predefined nucleation site for the skyrmion of interest. A movie of the series is available as Supporting Information. The skyrmion at the edge of the field of view is presumably moved out of the short racetrack (\SI{2}{\micro m} length) in the course of the motion to the left and then pinned in a region with lower current density. Scalebar corresponds to \SI{500}{nm}. (h) Distance travelled in $x$ and $y$-direction as indicated by the arrows.}
\end{figure*}

For many applications, the subsequent detachment of a skyrmion from its predefined nucleation site will be important. In the following, we will thus discuss the possibility for detachment and guided motion after deterministic nucleation. A first detachment experiment is presented in Fig.~\ref{fig:detachment}, performed via FTH-based imaging with a field of view (FOV) of \SI{1.2}{\micro m} diameter. We started by applying a single bipolar current pulse ($j_x = \SI{\pm 1.4e12}{A \per m^2}$) to nucleate skyrmions on a He$^+$-ion irradiated dot array with \SI{60}{nm} dot diameter. We first imaged the nucleation from saturation five times to demonstrate a reliable skyrmion formation in the irradiated dot. While the skyrmion nucleated at the central dot is of interest (Fig.~\ref{fig:detachment}a), a second skyrmion is nucleated at a neighboring dot visible in our FOV. In order to detach and move the central skyrmion via SOT to the left, we applied a sequence of single unipolar current pulses of reduced current density ($j_x = \SI{4.3e11}{A \per m^2}$) as compared to the nucleating bipolar pulse and recorded an image after every pulse (Fig.~\ref{fig:detachment}b--g). With the first unipolar pulse, the central skyrmion mostly deforms which we interpret as being part of the depinning from the nucleation site. Once detached, the skyrmion can move freely back and forth. After three unipolar current pulses (Fig.~\ref{fig:detachment}b--d), we reversed the current direction and the skyrmion moves in the opposite direction (Fig.~\ref{fig:detachment}e--g). Important in the context of device applications, the magnetic field is kept constant during the entire sequence including nucleation, detachment and motion. Tracking the position of the skyrmion in $x$- and $y$-direction (Fig.~\ref{fig:detachment}h), we observe almost uniform motion steps along the current direction with an average velocity of $v_x = \SI{11\pm 3}{m/s}$. The typical inclination of the motion path owing to the skyrmion Hall effect \cite{litzius2017skyrmion,jiang2017direct} directly proves the topological nature of the magnetic textures created---they are skyrmions. 

\begin{figure}[htbp]
	\begin{center}
		\includegraphics{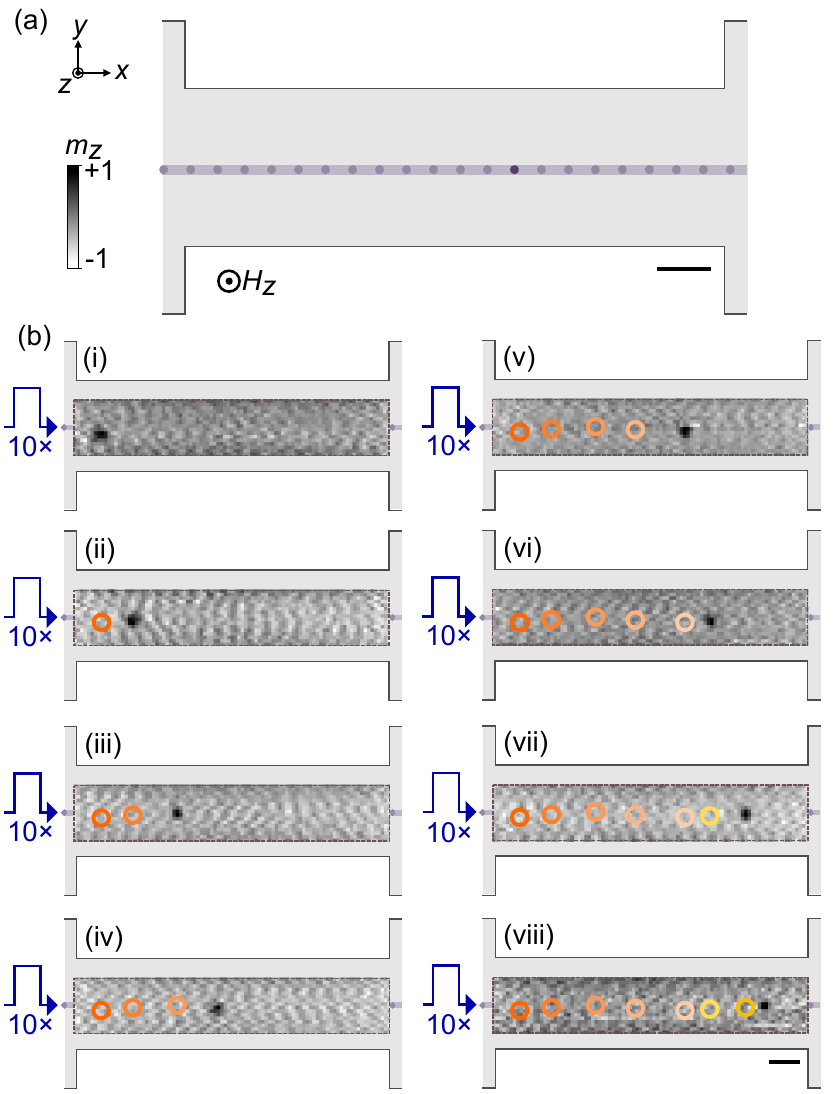}
	\end{center}
	\caption{\label{fig:guidedmotion}\textbf{Guided skyrmion motion.} (a) Schematic irradiation layout showing the nucleation dot (dark red), the guiding channel (light red) with intermediate medium-dose dots. (b) STXM images of a motion series at constant field of $\mu_0 H_z= \SI{39}{mT}$. After nucleation with a single bipolar pulse at the nucleation dot ($j_x = \SI{\pm 1.2e12}{A \per m^2}$), the skyrmion is imaged along the guiding chain in \SI{24} steps. Eight representative consecutive images are shown (i--viii). In order to increase the distance moved between two images, ten consecutive unipolar current pulses ($j_x = \SI{7.7e11}{A \per m^2}$) were applied before recording the next image. The skyrmion moves back and forth along the entire field of view with no systematic deviation from the irradiated channel, i.e., with suppressed skyrmion Hall effect. The orange-shaded circles mark previous positions. Scalebars correspond to \SI{500}{nm}.}
\end{figure}

Next, we address the need to guide the SOT-driven skyrmion motion within the racetrack in a fashion compatible with the deterministic nucleation and subsequent detachment demonstrated so far. To this end, we prepare a He$^+$ irradiation pattern of dots (\SI{60}{nm} diameter, \num{50} ions/nm$^2$), linked via \SI{250}{nm} long connecting lines (\num{25} ions/nm$^2$), as illustrated in Fig.~\ref{fig:guidedmotion}a. A single dot on the channel near the center of the racetrack was produced with a higher dose of \num{100} ions/nm$^2$ to act as a nucleation site. At this location, a single skyrmion nucleates from a single bipolar current pulse ($j_x = \SI{\pm 1.2e12}{A \per m^2}$) after saturation (not shown). Within the irradiated channel, we moved the skyrmion with unipolar current pulses ($j_x = \SI{7.7e11}{A \per m^2}$) at an average velocity of $v_x = \SI{2.5\pm 0.8}{m/s}$. We are able to propagate the skyrmion back and forth over several micrometers along a straight line, covering the entire length of the magnetic racetrack. For nucleation and SOT-driven motion, a constant magnetic field of \SI{39}{mT} was applied. Figure~\ref{fig:guidedmotion}b illustrates the guided motion with a selection of eight (i--viii) consecutive STXM images of the motion series. The full series including the nucleation at the central dot and all \num{24} motion steps is provided as movie in the Supporting Information. The skyrmion moves strictly along the guiding channel, deviations in $y$-direction are smaller than the skyrmion diameter of about \SI{125}{nm}. Note that the skyrmion also passes across its previous nucleation site without being trapped when moving back and forth. Our approach also directly supports the controlled synchronized motion of a train of skyrmions as we present in Fig.~S2 of the Supporting Information. The demonstration of this shift operation on a whole bit pattern represents a crucial step towards proposed memory applications of skyrmions\cite{fert2013skyrmions}. 

Our findings prove that He$^{+}$-ion irradiation represents an excellent tool to reliably and precisely localize current-induced and optical nucleation of isolated skyrmions in a magnetic multilayer material. On the one hand, the magnetic modification is effective enough to promote nucleation in irradiated areas with full reliability while nucleation in non-irradiated areas can be suppressed completely. On the other hand, the modification is gentle enough that the magnetic racetrack remains topographically intact and the magnetic properties in the irradiated regions are preserved in the sense that stable nanometer-scale skyrmions are formed, with a size and shape that we cannot distinguish from skyrmions in non-irradiated areas of the sample. The latter finding strongly suggests that a local pinning potential possibly created by ion irradiation can only play a minor role for the localization of the nucleation.

Instead, recent work reported that ion irradiation can significantly enhance the efficiency of SOT-driven magnetization switching in heterostructures containing a single ferromagnetic layer \cite{yun2019lowering, dunne2020helium, an2021improved}. The reasons were found in a reduced PMA of the irradiated structure \cite{yun2019lowering, dunne2020helium} and an increased spin-Hall angle from the Pt probably due to an increased defect density in this layer \cite{an2021improved}. In addition, intermixing at the Co-Pt interfaces can lead to a higher spin transparency \cite{zhu2019enhancement} and again to an increased spin-Hall angle \cite{zhang2022efficient}.  We expect that similar effects are key also for the easier SOT nucleation of skyrmions in He$^+$ irradiated areas of our material even though it is much thicker compared to material systems used in Refs.~\citenum{yun2019lowering, dunne2020helium, an2021improved}. Our SRIM simulations confirm a mostly homogeneous ion penetration along the depth direction of our multilayer (see Supporting Information). Consequently, the more efficient SOT switching allows us to act against a significantly higher applied field compared to the SOT in non-irradiated regions. In the case of laser-induced skyrmion nucleation, we also identified the reduced PMA in the He$^+$ irradiated areas as main origin for the preferred skyrmion nucleation. We required a reduced laser fluence to nucleate skyrmions in the ion-irradiated dots. We modelled this dependence in atomistic simulations of the high-temperature fluctuation phase which leads to the formation of skyrmions \cite{buttner2021observation} in model systems with different values of the PMA. For the low-PMA material, the simulations predict a lower excitation threshold for skyrmion nucleation (see Supporting Information). The laser fluence, thus, provides a handle to trigger the skyrmion nucleation exclusively in the predefined regions. 

In the second part of our work, we demonstrated the localized nucleation of a skyrmion combined with its detachment from the site and subsequent guided motion in a narrow ion-irradiated channel. Such a controlled detachment from the nucleation site and the unimpeded skyrmion motion across it are only possible if the pinning potential formed by the ion irradiation is sufficiently shallow and tunable---an important advantage of light-ions over the heavier ions such as Ga$^+$ that are routinely used for topographic modifications on the same length scale. Our method to create skyrmion nucleation sites in technologically relevant multilayers \cite{wiesendanger2016nanoscale,fert2017magnetic,jiang2017skyrmions} is directly compatible with previously proposed ion channels in single layers \cite{juge2021helium} to guide the skyrmion motion and suppress deflections due to the skyrmion Hall angle. Here, we demonstrate guided motion of a skyrmion over tens of micrometers distances back and forth in the magnetic racetrack. Yet, our measurements also indicate that the skyrmion velocity in the He$^+$ irradiated channel is significantly slower compared to the non-modified material, which contradicts predictions from simulations\cite{juge2021helium} and deserves more systematic investigation.  We consider our combination of nucleation and motion both guided by a He$^+$ created anisotropy landscape as a key development to foster controlled skyrmion manipulation on the nanometer scale. Depending on the thickness and composition of the magnetic material, even local material modification with a spatial extent in the single-digit nanometer range is possible, allowing generation and localization of ultrasmall skyrmions\cite{caretta2018fast} by He$^+$-ion patterning.

In summary, He$^+$ patterning can transform both SOT-driven and laser-induced skyrmion nucleation into a deterministic process and allows for subsequent guided motion by SOT-pulses. The approach demonstrated here and the achieved level of control may enable repetitive pump--probe experiments\cite{litzius2017skyrmion,woo2018deterministic} with precise real-space skyrmion localization on the nanometer scale and thus provide a platform for fundamental research on isolated skyrmions or other topological textures under controlled conditions. With respect to skyrmion computing applications, we achieved significant progress by demonstrating important operations in a device-compatible way. Based on minimally invasive modification of a racetrack, we gain control over single-skyrmion writing as well as shifting single and chained skyrmions. We expect that our approach is also compatible with SOT and laser-based skyrmion deletion methods\cite{woo2018deterministic,gerlinger2021application} and consecutive writing of skyrmion patterns within a track \cite{buttner2017field}. Our approach opens up a new route towards deterministic skyrmion logic and eventually may allow to reliably introduce defined skyrmion motion pathways in novel computing architectures \cite{luo2018reconfigurable,zazvorka2019thermal,zhang2020skyrmion,capic2020skyrmions,finocchio2021promise}.

\begin{suppinfo}
Extended experimental details on sample fabrication, excitation parameters, and imaging methods; Images of a racetrack taken with scanning electron microscopy, atomic force microscopy and STXM, demonstrating damage-free patterning; STXM image series demonstrating guided motion of a skyrmion train; Methodical details and results of the SRIM simulations and atomistic spin simulations, including figures on the simulations' results; Movies of the skyrmion motion presented in Figs.~\ref{fig:detachment} and \ref{fig:guidedmotion}.
\end{suppinfo}

\begin{acknowledgement}
Measurements were carried out at PETRA III (DESY) and BESSY II (Helmholtz-Zentrum Berlin, HZB). We acknowledge DESY (Hamburg, Germany), a member of the Helmholtz Association HGF, for the provision of experimental facilities at PETRA III, beamline P04. We thank Helmholtz-Zentrum Berlin for the allocation of synchrotron radiation beamtime. Ion beam patterning was performed in the Corelab Correlative Microscopy and Spectroscopy at HZB. Financial support from the Leibniz Association via Grant No. K162/2018 (OptiSPIN) and the Helmholtz Young Investigator Group Program is acknowledged. Furthermore, we would like to highlight the support from the EU COST Action CA 19140 (FIT4NANO). This work is part of the Shell-NWO/FOM-initiative “Computational sciences for energy research” of Shell and Chemical Sciences, Earth and Life Sciences, Physical Sciences, FOM and STW, and received funding from the European Research Council ERC grant agreement No. 856538 (3D-MAGiC).
\end{acknowledgement}

\bibliography{bibliography}

\newpage

\end{document}